\documentclass[12pt]{iopart}
\usepackage{iopams}

\usepackage{graphicx}
\usepackage{slashed}

\begin{document}
\title{Transport properties of the Coulomb-Majorana junction}
\author{Alex Zazunov$^1$, Alexander Altland$^{2}$ and Reinhold Egger$^1$}
\address{$^1$~Institut f\"ur Theoretische Physik, Heinrich-Heine-Universit\"at,
D-40225 D\"usseldorf, Germany}
\address{$^2$~Institut f\"ur Theoretische Physik, Universit\"at zu K\"oln,
Z\"ulpicher Str.~77, D-50937 K\"oln}
\ead{egger@thphy.uni-duesseldorf.de}
\begin{abstract}
We provide a comprehensive theoretical description of low-energy
quantum transport for a Coulomb-Majorana junction, where several
helical Luttinger liquid nanowires are coupled to a 
joint mesoscopic superconductor
with finite charging energy.  Including the Majorana bound states 
formed near the ends of superconducting wire parts,
we derive and analyze the Keldysh phase action describing
nonequilibrium charge transport properties of the junction.
The low-energy physics corresponds to a two-channel Kondo model with 
symmetry group $\mathrm{SO}(M)$, where $M$ is the number of leads
connected to the superconductor.  Transport observables,
such as the conductance tensor or current noise correlations,
display non-trivial temperature or voltage dependences reflecting
non-Fermi liquid behavior.
\end{abstract}


\section{Introduction}

The quantum transport properties of topological insulators and
topological superconductors have attracted a lot of recent interest
\cite{hasan,qizhang}.  One prominent example concerns the localized
Majorana bound states (MBSs) forming at the boundaries of
one-dimensional topological superconductor wires. Thanks to their
non-Abelian statistics, these exotic states, once realized
successfully, might become useful in topological quantum computation
applications \cite{beenakker,alicea,karsten}.  Majorana nanowires have
been proposed for several material platforms \cite{alicea}, including
semiconductor (InSb or InAs) nanowires with strong spin-orbit
coupling, where the topological phase is realized in a Zeeman field by
proximity coupling to a conventional $s$-wave BCS
superconductor~\cite{lutchyn,oreg}.  Once such a nanowire is contacted
to a normal metal electrode, the MBS builds up a zero energy
resonance, which in turn causes a resonant Andreev reflection
conductance peak in the tunneling
conductance~\cite{demler,sodano,nilsson,law,flensberg,wimmer,chung,golub,bose,fidkowski,lutchyn1}.
Signatures of this type have been observed
experimentally~\cite{leo,exp1,exp2,exp3,exp4}, although an unambiguous
identification as Majorana bound states is pending.  

Here we study the possibility of realizing and observing novel quantum
transport phenomena caused by \textit{Coulomb interactions} in
Majorana devices, including non-Fermi liquid behavior.  We analyze the
'Coulomb-Majorana junction' schematically shown in Fig.~\ref{fig1},
where a floating (not grounded) mesoscopic superconductor is
responsible for the proximity-induced pairing in several $(N)$
Majorana nanowires.  The nanowire parts not in contact to the
superconductor serve as normal-conducting leads (as in the experiments
of Ref.~\cite{leo}), and we have $M\le 2N$ normal leads.  Near each
boundary of a given superconducting wire part, we assume the existence
of a MBS, see Fig.~\ref{fig1}.  In total, we then have $2N$ Majorana
fermions on the central superconducting island ('dot'). 
The dominant coupling between the dot and the $j$th
lead involves tunneling through the respective MBS with coupling
strength $t_j$.  Additional coupling mechanisms turn out to be
irrelevant on energy scales below the proximity-induced gap
\cite{fidkowski}, which is the regime of interest here.  In the
absence of Coulomb interactions, the standard resonant Andreev
reflection picture applies where currents flowing through different
leads are completely decoupled \cite{demler}. This decoupling includes
noise correlations and all higher-order cumulants.

Coulomb interactions now play a two-fold role in this 
system.  First, for each of the $M$ nanowire parts representing
a 'lead electrode' (without pairing but including the Zeeman field 
and spin-orbit coupling), interactions imply that we are dealing with an 
effectively spinless helical Luttinger liquid (hLL).
The hLL is characterized by a dimensionless interaction parameter
$g\le 1$ \cite{alex2,gogolin,delft}, where $g=1$ corresponds to 
the non-interacting limit.  Second, we also have on-dot Coulomb interactions. 
Several works have already shown that MBSs survive the presence
of weak repulsive electron-electron interactions in the superconducting 
nanowire \cite{loss1,stoud,eran}. However, these interactions also introduce
correlations between the Majoranas and thereby entangle
different connecting leads for a device as shown in Fig.~\ref{fig1}.
Here we shall focus on the universal regime of 
long (compared to the typical MBS size) and well-separated Majorana
wires, such that all direct tunneling couplings connecting the Majoranas can 
be neglected, and only the charging energy of 
the dot, $E_c$, generates inter-wire couplings.
Since Coulomb charging effects are often tunable by gate voltages,
this option could be attractive for braiding protocols 
in $Y$ or $X$ junctions of Majorana nanowires, which
 so far have been based on direct tunneling contacts
 \cite{alicea1,halperin,zazu1}.  For $M=2$, our model gives 
the Majorana single-charge transistor~\cite{fu,xu,zazu,heck2,msct,heck}
which features, for instance, a universal halving of the peak conductance 
with increasing $E_c$.  Even more remarkable effects 
are predicted for $M>2$ terminals, where  
the resonant Andreev reflection fixed point is unstable against 
interactions and non-Fermi liquid behavior due to a
topological $\mathrm{SO}(M)$ Kondo effect is 
expected \cite{beri1,tsvelik,cmj,beri2,crampe}.
We note that only the $M$ MBSs tunnel-coupled to lead electrodes 
affect our final results, while the remaining $(2N-M)$ MBSs act as
'spectator' modes.  Throughout this paper, we assume that $E_c$
is sufficiently strong to allow for charge quantization effects on the
island.

To set the stage for our subsequent discussion, we now summarize the
picture emerging from an effective phase action approach to 
the interacting problem \cite{cmj}.  An intuitive interpretation of 
tunneling processes from (into) lead $j$ follows by viewing these
as  'particles' (antiparticles) with flavor index $j$. 
At high effective energy scales, $\omega\gtrsim E_c$,
such particles are 'asymptotically free' in the sense that the 
tunneling amplitudes $t_j$ independently scale upwards when lowering 
the scale $\omega$ during the renormalization group (RG) flow.
This increase of the $t_j$ reflects a flow towards the putative
resonant Andreev reflection fixed point. However, 
this RG flow will be stopped by 'confinement' when reaching the energy scale
$\omega \sim E_c$, where electroneutrality 
enforces that in-tunneling events must be followed by successive out-tunneling
events. For $\omega\lesssim E_c$, the theory is then best expressed in 
terms of 'dipoles' (strongly bound particle-antiparticle pairs)
corresponding to almost instantaneous charge transmission from 
lead $j$ to lead $k \not=j$.
The effective dipole coupling strengths, $\lambda_{jk}$, are subject
to downward renormalization due to the well-known
suppression of the hLL tunneling density of states \cite{gogolin}, 
and upward renormalization due to dipole-dipole fusion events.
For $M>2$ interacting leads ($g<1$), this competition results in an isotropic 
repulsive fixed point, $\lambda^\ast$, separating a flow
towards the decoupled dot ($\lambda\to 0$) from a flow towards an
exotic Kondo regime ($\lambda\to \infty$).  It turns out that for not too
large $E_c$, the low-energy RG flow always proceeds towards the 
strong-coupling topological Kondo regime.  
As we will explain below, this corresponds
to an isotropic two-channel Kondo effect with the orthogonal symmetry 
group $\mathrm{SO}(M)$, which emerges on energy scales $\omega<T_K$  
below the Kondo temperature $T_K$ defined in Eq.~(\ref{tk}) below. 
This fixed point exhibits local non-Fermi liquid behavior and
is always reached for non-interacting ($g=1$) leads.

The above physics naturally determines the temperature or voltage 
dependence of typical quantum transport 
observables such as the conductance tensor, $G_{jk}$, or the
current noise correlations, $S_{jk}$, defined in Eqs.~(\ref{conddef}) and 
(\ref{noisedef}), respectively.  
In particular, the voltage-dependent \textit{shot noise} \cite{blanter} 
encoded in $S_{jk}$ may provide valuable information about
two-particle entanglement and nonlocality not contained in the conductance.
For small transmitted or backscattered current $I$, it is customary to 
define the \textit{Fano factor}, $F=S/eI$, comparing the shot noise to 
its Poissonian reference value.  For conventional Coulomb-blockaded 
spin-degenerate quantum dots, shot noise 
in the sequential tunneling regime is generally sub-Poissonian, 
$F\le 1$, while cotunneling allows for super-Poissonian  
noise \cite{loss2,thielmann}.  At energies below $E_c$, 
a single-channel Kondo effect with symmetry group $\mathrm{SU}(2)$ 
can be realized in such a setting, where  
for voltages above the Kondo temperature,
$V>T_K$, shot noise shows logarithmic scaling, $S\sim \ln^{-2}(V/T_K)$, 
with a peak around $V\sim T_K$ \cite{meir}.
For $V\ll T_K$, one finds shot noise suppression, $S\sim V^3$, from 
a local Fermi liquid approach \cite{sela,komnik,golub2}, 
implying the universal Fano factor $F=5/3$ also observed 
experimentally \cite{kexp,kexp2}.
(Intra-lead interactions, $g<1$, weakly affect this result
 \cite{golub3}.) When additional orbital degeneracies are present, 
an $\mathrm{SU}(N)$ variant of this scenario can be realized, 
where local Fermi liquid theory holds again \cite{mora1,mora2,sakano1,sakano2}. 
The Fano factor remains universal (but different from $5/3$) 
in the $\mathrm{SU}(N)$ Kondo regime, cf.~Table I in Ref.~\cite{mora2}. 
Experimental studies of shot noise for $\mathrm{SU}(4)$ Kondo dots 
have also been reported \cite{sasaki,delattre}.  
Finally, a two-channel $\mathrm{SU}(2)$ Kondo effect was
observed in Ref.~\cite{twochannel}. However, 
the energy dependence of transport observables is expected to
differ from the two-channel $\mathrm{SO}(M)$ case at hand.

The structure of the remainder of article is as follows.
In Sec.~\ref{sec2}, the model is described and the
phase action determining the Keldysh generating functional  
will be derived. The latter gives access to the full counting 
statistics of charge transport in this system.  
In Sec.~\ref{sec3}, we consider the theory on energy scales
below $E_c$.  We discuss in detail the connection of the 
phase action approach to the topological
Kondo effect, including a derivation of the dual 'instanton' action
capturing the physics below the Kondo temperature.  
Our results for the differential conductance and, in particular,
for the shot noise tensor, are presented in Sec.~\ref{sec4}, 
followed by concluding remarks in Sec.~\ref{sec5}. 
Technical details can be found in 
the Appendix, and we often use units with $e=\hbar=k_B=1$.

\section{Model and Keldysh phase action}\label{sec2}

\begin{figure}[t]
\begin{center}
\includegraphics[width=0.7\textwidth]{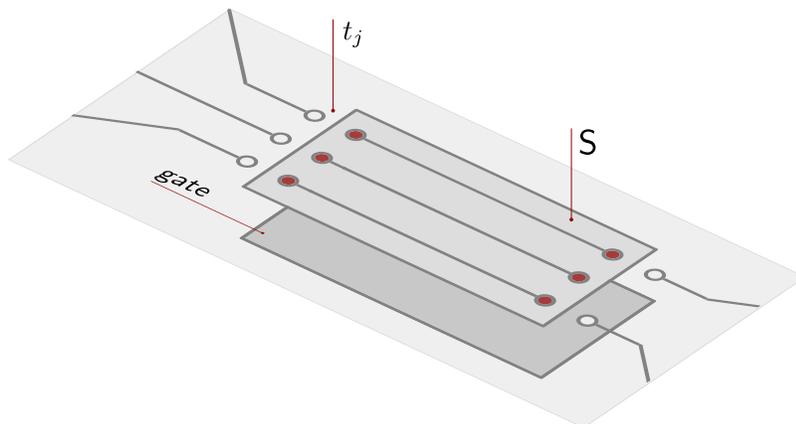}
\end{center}
\caption{\label{fig1} 
Multiterminal Coulomb-Majorana junction setup (schematic):
$N$ helical nanowires (here $N=3$) on top of a floating 
mesoscopic superconductor film with charging energy $E_c$
are assumed to host Majorana bound states. A backgate electrode
allows to gate the superconductor.  Each wire holds
a pair of MBSs (indicated as filled circles)
associated with Majorana fermion operators $\gamma_j$
near the ends of proximity-coupled 
nanowire parts.  The $M\le 2N$ normal-conducting wire segments
away from the dot (here $M=5$) act as helical Luttinger liquid leads, 
tunnel-coupled to the dot with amplitudes $t_j$.  
Klein factors are expressed in terms of Majorana fermions $\eta_j$
(open circles).  For the case of $M<2N$ leads, 
we put one or several $t_j=0$.  }
\end{figure}

Consider the multiterminal Coulomb-Majorana junction
with $M$ connecting leads schematically shown in Fig.~\ref{fig1}.
We start by introducing an appropriate Hamiltonian describing
this system on energy scales below the proximity-induced  
superconducting gap in the nanowires. 

\subsection{Low-energy model: Hamiltonian}

The Hamiltonian is written as $H=H_c+H_t+H_l$, with the dot
Hamiltonian $H_c$, the tunneling Hamiltonian $H_t$, and $H_l$ for the
normal-conducting hLL leads.  Labeling the different nanowires by
$\alpha=1,\ldots,N$, for each wire we assume that two spatially well
separated MBSs are present, corresponding to the Majorana fermion
operators $\gamma_{2\alpha-1}$ and $\gamma_{2\alpha}$, where
$\gamma_j=\gamma_j^\dagger$ with $\{\gamma_j,\gamma_{k}\}
=\delta_{jk}$.  It is convenient to define nonlocal auxiliary fermion
operators $d_\alpha= (\gamma_{2\alpha-1}+i\gamma_{2\alpha})/\sqrt{2}$,
with total number operator $\hat n=\sum_\alpha d^\dagger_\alpha
d^{}_\alpha$.  For the parameter regime of interest, with the
proximity-induced gap constituting the largest energy scale,
quasiparticle excitations in the superconductor can be neglected.
Hence the Majorana fermions, $\gamma_j$, and the Cooper pair number
operator, $\hat N_c$, are the only important dot degrees of freedom.
Note that $\hat N_c$ is conjugate to the condensate phase $\varphi$,
i.e., we have $[\varphi,2\hat N_c]=i$ and the operator $e^{- 2 i
  \varphi}$ annihilates a Cooper pair, $N_c\to N_c-1$. Since at this
stage all dot variables are zero-energy modes, the dot Hamiltonian
$H_c$ is fully expressed by the Coulomb charging term,
\begin{equation}\label{hc}
H_c = E_c (2\hat N_c+\hat n-n_g)^2,
\end{equation}
where the dimensionless offset charge $n_g$ can be continuously varied
by a background gate voltage.  Next the semi-infinite $(x>0)$ hLL
leads, with tunneling contacts connecting the respective lead to the
dot at $x=0$, are described by dual pairs of bosonic fields,
$\phi_j(x)$ and $\theta_j(x)$, with the hLL Hamiltonian
\cite{fidkowski,gogolin}
\begin{equation}\label{hl}
H_l =\frac{v}{2\pi} \sum_{j=1}^M \int_0^\infty dx \left[
g(\partial_x\phi_j)^2+ g^{-1}(\partial_x\theta_j)^2\right].
\end{equation}
For simplicity, we assume identical Fermi velocity $v$ and hLL
parameter $g$ for all wires, with weakly repulsive interactions such
that $1/2<g\le 1$.  The bosonized right- or left-moving fermion
annihilation operator reads \cite{gogolin} $\psi_{j,R/L}(x)= a^{-1/2}
\eta_j e^{i[\phi_j \pm \theta_j](x)}$, where $a$ is a short distance
cutoff.  We have also introduced a set of auxiliary Majorana fermions
$\eta_j$, with $\{\eta_j, \eta_{k}\}=\delta_{jk}$, to represent the
'Klein factors' \cite{delft,oshikawa} enforcing fermion
anticommutation relations between different leads.  To ensure open
boundary conditions at $x=0$ in the absence of tunneling, we require
$\psi_{j,L}(0)=\psi_{j,R}(0)$, thereby pinning all boson fields
$\theta_j(0)=0$.  The lead fermion operators near $x=0$ are thus written
as $\Psi_j= a^{-1/2}\ \eta_j\ e^{i\phi_j(x=0)}$.  Finally, as derived
in Refs.~\cite{fu,zazu}, the tunneling Hamiltonian connecting leads and
dot reads
\begin{equation}\label{ht}
H_t = \sqrt{a/2} \sum_j  t_j
\Psi_j^\dagger \left( d^{}_{\alpha_j} + (-)^{j-1} e^{-2i\varphi} 
d_{\alpha_j}^\dagger \right) + {\rm H.c.},
\end{equation}
where $\alpha_j=[j/2]+1$.  The term $\propto \Psi^\dagger d$
describes the  transfer of a fermion from the dot to the lead by 
annihilation of a
$d$-fermion.  The term $\propto \Psi^\dagger e^{-2i\varphi} d^\dagger$
represents an alternative way of annihilating a dot fermion, viz. by
\textit{creation} of a Majorana $d$-fermion along with annihilation of
a Cooper pair.  Without loss of generality, the bare tunneling
amplitudes are taken real and positive, $t_j>0$.  Using Eq.~(\ref{ht})
and the Heisenberg equation of motion, the operator
\begin{equation}\label{currop}
\hat I_j=  i t_j \sqrt{a/2} \
\Psi_j^\dagger \left( d^{}_{\alpha_j} + (-)^{j-1} e^{-2i\varphi} 
d_{\alpha_j}^\dagger \right) + {\rm H.c.}
\end{equation}
describes the current flowing from the $j$th lead towards the dot.

\subsection{Real-time phase action}\label{sec2b}

We next derive an action $S$ representing the above 
second-quantized Hamiltonian. In anticipation of our later 
application of the Keldysh formalism, we consider a real-time version
of the theory with the action $S=S_c+S_l+S_t+S_f$, 
\begin{eqnarray}\nonumber 
S_c[N_c,\varphi,d,\bar d] &= \int dt\,
\left[ 2N_c \dot \varphi-H_c(N_c,d,\bar d) \right],\\ \nonumber
S_l[\theta,\phi]&= \frac{1}{\pi}\sum_j\int
dx dt \,\partial_t \phi_j \partial_x \theta_j -\int dt\,
H_l[\phi,\theta],\\ \nonumber
S_t[d,\bar d,\eta,\phi,\varphi] &= - \int dt\, H_t(d,\bar
d,\eta,\phi(x=0),\varphi),\\ \label{fullact}
S_f[d,\bar d,\eta]&=i\int dt\, \left(\sum_\alpha \bar d_\alpha \dot 
d_\alpha +\frac12 \sum_j \eta_j\dot\eta_j\right),
\end{eqnarray}
where $N_c,\phi,d,\bar d,\dots$ are the real or Grassmann valued field
variables corresponding to the operators $\hat N_c, \hat \phi,
d,d^\dagger,\dots$ \cite{alex2}.  
We start out by integrating over all 
those variables of the theory which do not enter in a 
non-trivial (non-quadratic) form. The fact that the tunneling 
operator couples only to the field amplitudes
$\phi$ suggests to integrate over the conjugate fields $\theta_j(x,t)$, 
which yields the $\phi$-representation of the hLL action,
\begin{equation} \label{eq:4}
 S_l[\phi]={g\over 2\pi v}\sum_j\int dx dt\,\phi_j\left(-\partial^2_t
 +v^2 \partial_x^2\right)\phi_j.
\end{equation}
Similarly, integration over $N_c$ brings the charging action $S_c$ 
into the form 
\begin{equation}\label{sc2}
S_c[\varphi,d,\bar d]= -2\pi  W n_g+ 
\int dt \left( \frac{\dot\varphi^2}{4E_c} 
-\sum_\alpha \dot\varphi\ \bar d_\alpha d_\alpha \right),
\end{equation}
where the presence of the integer-valued winding number $W$
reflects the discreteness of the variable $N_c$. 
The summation over the winding number $W$ encodes  charge quantization
due to the charging energy.  We below consider only
$n_g$ values close to an integer, where the charge state of the dot
is well defined.  As we show in \ref{appa}, 
the $W$ summation can then effectively be discarded.

We next remove the term $\sim \dot \varphi \bar d_\alpha
d_\alpha$ by the gauge transformation $d_\alpha \to e^{-i\varphi}
d_\alpha$. A side effect of this transformation is that the tunneling
action $S_t$ now assumes a more symmetric form,
\begin{equation}\label{tunnel}
S_t[\gamma,\eta,\Phi,\varphi]= \sum_j t_j \int dt\,\hat\sigma_j 
\sin(\Phi_j+\varphi),
\end{equation}
where we introduced the notation $\Phi_j(t)=\phi_j(x=0,t)$. 
In addition, we turned back to dot-Majorana fields,
$\gamma_{2\alpha-1}= (d_\alpha^{} + d_\alpha^\dagger)/\sqrt{2}$ and
$\gamma_{2\alpha}=-i(d_\alpha^{}-d^\dagger_{\alpha})/\sqrt{2}$, and
defined $\hat\sigma_j\equiv 2i\gamma_j\eta_j$. 
In essence, the Majorana fermions have been removed from  
the charging energy (\ref{hc}) through this gauge transformation,
and now couple to $\varphi$ only through the tunneling term
(\ref{tunnel}). Since our model assumes all direct tunneling matrix
elements between different MBSs to vanish, the Majorana fermions
only appear through the $M$ operators $\hat\sigma_j$.
By construction, these operators do (i) commute with 
the system Hamiltonian, $[H,\hat\sigma_j]=0$, (ii) square to unity,
$\hat\sigma_j^2=1$, and (iii) mutually commute, $[\hat \sigma_j,\hat
\sigma_k]=0$.  According to (iii), all operators 
$\hat \sigma_j$ can be diagonalized simultaneously. According to (i) and
(ii), the two possible eigenvalues $\sigma_j=\pm 1$ are dynamically conserved. 
Instead of working with the operators $\hat\sigma_j$ 
and the Grassmann action piece $S_f$ in Eq.~(\ref{fullact}) explicitly,
we may therefore multiply each tunneling amplitude $\sim
\sin(\Phi_j+\varphi)$ with an independent sign factor $\sigma_j=\pm 1$ 
and then sum over these.  

Next, we note that a uniform shift, $\Phi_j(t)\to \Phi_j(t)+\pi$, only
changes the sign of the respective tunneling term but leaves the remaining
action invariant.  The sign factor $\sigma_j$ can thereby be gauged 
away, and the perturbation series in $S_t$ will
automatically contain only even orders in $t_j$.
The above reasoning allows us to ignore all Grassmann fields as well as the
sign factors $\sigma_j$. 
It is worth noting that this is an enormous simplification compared to 
other multiple Luttinger liquid tunneling contexts 
\cite{oshikawa,nayak,chen,alex}, where the presence of Klein
factors ($\eta_j$) leads to complicated correlations. 
To summarize the above steps, the effective real-time action 
is given by $S=S_c+S_l+S_t$, where 
\begin{equation}
 S_c[\varphi]= -\frac{1}{4E_c}  \int dt\, \varphi \ddot \varphi,
\end{equation}
$S_l$ is defined in Eq.~(\ref{eq:4}), and $S_t$ in Eq.~(\ref{tunnel}).
Notice that we are left with an action involving the phase-like fields
$\varphi(t)$ and $\phi_j(x,t)$ only.

\subsection{Keldysh generating functional and 
transport observables} \label{sec2c}

In the next step, we put the real-time phase action $S$ onto a Keldysh
contour and couple it to source fields, $\chi_j$, suitable for the calculation
of transport observables \cite{alex2}.  To this end, let us imagine
that the system can be described by an initial density matrix
$\rho_{0}$ at time $t=-t_0/2$, where tunneling between
leads and dot is assumed absent.  Each of the $M$ leads thus has its
own grand-canonical equilibrium density matrix with chemical potential
$\mu_j$.  We here do not discuss thermal transport and thus assume
identical temperature $T$ for all leads.  The initial state $\rho_0$
is then time-evolved under the full Hamiltonian $H$ (including
tunneling) along the forward $(+)$ part of the Keldysh time contour up
to $t=t_0/2$, followed by backward $(-)$ time evolution all the way
back to $t=-t_0/2$.  Eventually, the limit $t_0\to \infty$ will be
taken.  Using standard notation \cite{alex2}, we introduce
time-dependent counting fields $\chi_j(t)$, probing the fluctuating
current from the $j$th wire to the dot at time $t$.  In terms of the
dynamical lead fermion fields on the two Keldysh branches,
$\psi_{j,\pm}(x,t)$, we gauge out the chemical potentials $\mu_j$ and
include the counting fields $\chi_j$ as phase factors,
$\psi_{j,\pm}(t)\to e^{i[\mu_j t \pm \chi_j(t)/2]} \psi_{j,\pm}(t),$
which therefore appear solely in the tunneling term.  Note that the
counting fields appear with opposite signs on the forward and backward
parts of the Keldysh contour.  The resulting Keldysh generating functional,
with normalization $Z[0]=1$ and time-ordering operator ${\cal T}_K$ along
the Keldysh contour, 
\begin{equation}\label{fcs1}
Z[\chi] =  \frac{{\rm Tr} \left( {\cal T}_K 
e^{+i H_{-\chi} t_0}  \rho_0  e^{-i H_{+\chi} t_0} 
\right) }{ {\rm Tr}\rho_0},
\end{equation}
then encodes the complete information about charge transport 
statistics in our device.  

Expectation values involving the current operators $\hat I_j$ in
Eq.~(\ref{currop}) follow as functional derivatives of $Z[\chi]$ with
respect to the counting fields.  For instance, the mean current
flowing through the $j$th contact is given by
\begin{equation}\label{currentdef}
I_j(t) \equiv \langle \hat I_j(t)\rangle= 
-i\left. \frac{\delta \ln Z[\chi]} {\delta \chi_j(t)} \right|_{\chi=0}.
\end{equation}
Under steady-state conditions, $I_j$ is time independent and 
we may define the multiterminal differential conductance tensor,
\begin{equation}\label{conddef}
G_{jk} (\{\mu_i\}) \equiv - e \frac{ \partial I_j}{\partial \mu_k}.
\end{equation}
The temperature dependent linear conductance tensor $G_{jk}(T)$ then follows from Eq.~(\ref{conddef})
in the near-equilibrium regime ${\rm max}|\mu_j-\mu_k|\ll T$. Similarly, 
the symmetrized current noise correlations are contained in \cite{blanter}
\begin{equation} \label{noisedef}
S_{jk}(t-t') \equiv  \frac12 \left \langle\left [ 
\Delta \hat I_j(t), \Delta \hat I_k(t') \right ]_+ \right \rangle 
= - \left.\frac{\delta^2 \ln Z[\chi] }
{\delta \chi_j(t)\delta \chi_k(t')}\right|_{\chi=0},
\end{equation}
with the current fluctuation operators $\Delta \hat I_j\equiv \hat I_j-I_j$.
Under steady state conditions, $S_{jk}$ depends only on the time
difference $t-t'$, and we switch to the Fourier-transformed
noise tensor, $S_{jk}(\omega)$.  Near thermal equilibrium, 
the linear conductance matrix $G_{jk}$ and the Johnson-Nyquist 
noise tensor $S_{jk}$ are linked by the fluctuation dissipation 
theorem \cite{alex2}, $S_{jk}(\omega) = \omega \coth(\omega/2T) G_{jk}$.
Likewise all higher-order cumulants can in principle be extracted from 
the Keldysh functional (\ref{fcs1}).  Fluctuation relations
then impose symmetry relations for $Z[\chi]$ and thereby 
allow to generalize the fluctuation dissipation theorem 
to the nonequilibrium case.  This implies a relation connecting 
the third cumulant and shot noise, cf.~Ref.~\cite{fluctthm} and
references therein.

\subsection{Keldysh phase action} \label{sec2d}

Following the steps in Sec.~\ref{sec2b}, we now represent
$Z[\chi]$ as a functional integral over phase fields, 
$\varphi_s(t)$ and $\phi_{j,s}(x,t)$,  for the two 
Keldysh contour parts $s=\pm$. The extension of the phase action 
$S=S_c+S_l+S_t$ to the Keldysh theory reads as
\begin{eqnarray*}
 S_c[\varphi]&= -\frac{1}{E_c}  \int dt\, \varphi_q \ddot\varphi_c,
\cr
  S_l[\phi]&={2g\over \pi v}\sum_j\int dx dt\,\phi_{j,q} \left(-\partial^2_t
 +v^2 \partial^2_x\right)\phi_{j,c},\cr
S_t[\Phi,\varphi]&=
\sum_{s=\pm}\sum_j st_j \int dt\, \sin\left(\Phi_{j,s}+\varphi_s+\mu_j
  t+s\chi_{j}/2\right),
\end{eqnarray*}
where $\xi_c=(\xi_++\xi_-)/2$ and $\xi_q= (\xi_+-\xi_-)/2$
denote the classical and quantum components, respectively, of the
field variables $\xi=(\varphi,\phi_j)$.  The Fermi distribution functions
controlling the thermal occupation of lead modes at 
the initial time $t=-t_0/2$ are implicit in our notation. 

Next we integrate over the Gaussian fluctuations of field modes away
from the junction, $\phi(x\not=0)$. After a Fourier transformation, 
$S_l[\phi]$ gets thereby reduced to the action
\begin{equation} \label{eq:8}
 S_l[\Phi]=\frac12 \sum_j \int {d\omega\over 2\pi}
  \Phi_j^T(-\omega)G^{-1}(\omega)\Phi_j(\omega),
\end{equation}
with the Keldysh vector $\Phi\equiv \phi(0)= (\Phi_c,\Phi_q)^T$ containing 
the lead phase fields at $x=0$.  The dissipative Green's function matrix in 
Keldysh space is
\begin{equation}\label{gf}
 G= \left( \begin{array}{cc} G_K & G^+ \\ G^- & 0 \end{array} \right),
\qquad G^{\pm}(\omega)=\mp{i\pi\over 2g \omega},
\end{equation}
where the Keldysh component is given by
$G_K(\omega)=(G^+-G^-)(\omega)\coth ( \omega/ 2T)$.
The action $S_l[\Phi]$ describes Ohmic dissipation
generated by the 'bath' of lead modes kept at temperature $T$. 

Finally, we remove $\varphi$ from the tunneling term 
by a shift $\Phi_j \to \Phi_j -
\varphi$. This generates a linear coupling $\sim \varphi \Phi_0$
from Eq.~(\ref{eq:8}), where $\Phi_0 \equiv {1\over \sqrt M} \sum_j \Phi_j$.
Since $\varphi$ couples only to the 'zero mode' $\Phi_0$, it 
is beneficial to represent the field vector $\Phi\equiv \{\Phi_j\}$ as
\begin{equation}\label{alphadef}
 \Phi = \Phi_0  e_0 + \sum_{i=1}^{M-1} \alpha_i \  (e_{i+1}-e_i),
\end{equation}
with $e_i=(0,\dots,1,\dots,0)$ denoting a standard basis vector in
lead channel space and $e_0\equiv {1\over \sqrt M}(1,\dots,1)$. The
$M-1$ time-dependent fields $\alpha_i=(\alpha_{i,c},\alpha_{i,q})^T$
span the orthogonal complement of the zero mode $\Phi_0$.  The
overlap, $\Phi^{T}\Phi=\Phi_{0}^T\Phi_{0}-\alpha^{T}\Delta \alpha$,
between the new basis vectors is described by the $(M-1)\times (M-1)$
matrix
\begin{equation} \label{eq:3}
\Delta= -\left( \begin{array}{ccccc} 2&-1&0&\dots&\cr
      -1&2&-1&\dots&\cr &\ddots&\ddots&\ddots&\cr &\dots&-1&2&-1\cr
      &\dots&0&-1&2 \end{array} \right).
\end{equation}
The subsequent integration over $\varphi$ then transforms the
Keldysh action into $S[\Phi]=S_0+S_t$, with
\begin{eqnarray} \label{eq:7}
S_0&=& {1\over 2}\int {d\omega\over 2\pi}
\left[  \Phi_0^T(-\omega)G_0^{-1}(\omega)\Phi_0(\omega)
  -\alpha^T(-\omega)  G^{-1}(\omega)\otimes \Delta
  \alpha(\omega) \right] ,\\ \nonumber
S_t&=& \sum_{s=\pm}\sum_{j=1}^M s t_j \int dt\, \sin\left(\Phi_{j,s}(t)+\mu_j
  t+s\chi_{j}(t)/ 2\right),
\end{eqnarray}
involving only the lead boson fields at $x=0$, see Eq.~(\ref{alphadef}).
The zero mode Green's function $G_0$ follows from
\begin{equation}\label{epsdef}
  G^\pm_0(\omega)=\mp{i\pi\left(1\pm i\epsilon/ \omega\right)
    \over 2g \omega}, \qquad \epsilon={2gM E_c\over \pi}.
\end{equation}
The physical meaning of Eq.~(\ref{epsdef}) is that at low energies,
$|\omega|\ll \epsilon\sim E_c$, the fluctuations of the 
zero mode $\Phi_0$ become free, $G^\pm_0(\omega)\sim 1/\omega^{2}$, 
as a consequence of the pinning of the conjugate charge
fluctuations.  Note that for $E_c=0$, all $\Phi_j$ phase fields
fluctuate independently, consistent with the completely decoupled
leads in the resonant Andreev reflection picture \cite{demler}.  

\section{Topological Kondo effect}\label{sec3}

In this section, the focus will be mostly on the scaling properties of the
unperturbed system, and hence we set $\mu_j=\chi_j=0$ throughout.
The chemical potentials and the counting fields will be restored
in Sec.~\ref{sec4} when addressing transport observables.
In a first step, we derive a Keldysh phase action 
describing the physics on energy scales below the charging energy.

\subsection{Low-energy Keldysh phase action}

Consider a perturbative expansion of $Z$ in the tunneling
couplings $t_j$, with the effective Keldysh phase action~(\ref{eq:7}).  
Adopting a Coulomb gas picture, we interpret the 'scattering operators'
${\cal O}_{j,s}^{\pm} (t) = e^{\pm i\Phi_{j,s}(t)}$ 
as particles ('quarks') and antiparticles living on the time axis.  
Each particle carries a 'flavor' index $j=1,\ldots,M$, the  Keldysh contour 
index $s=\pm$, and has the coupling constant ('charge') $-it_j/2$.
To study the properties of this interacting
particle gas, we employ standard RG methods \cite{alex2}.  In a given
RG step, all 'fast' $\Phi(\omega)$ modes within the energy shell
$\Lambda/b<|\omega|<\Lambda$ are integrated out, with rescaling
parameter $b>1$ and the high-energy cutoff $\Lambda$ initially given
by the proximity gap.  At the end of the RG step, we rescale all
energies, $\omega\to b\omega$, and thus $\Lambda$ remains invariant.
In a first stage of the RG analysis, we follow the RG flow 
by subsequently integrating over all modes from $\omega=\Lambda$
down to $\omega= E_c$ \cite{foot}.
For small $t_j$, the particle density is low and different $t_j$ 
renormalize independently.  Noting from Eqs.~(\ref{eq:7}) 
and (\ref{epsdef}) that for $\omega>E_c$, 
the zero mode $\Phi_0$ stays basically unaffected by 
the charging energy, the $t_j$ are relevant scaling fields
with net scaling dimension $1-1/2g$  \cite{fidkowski}. 
Once the RG flow has reached the energy scale $\omega=E_c$,
the renormalized tunneling couplings are given by
\begin{equation}\label{t11}
t_j^{(1)} = t_j (\Lambda/E_c)^{1-\frac{1}{2g}} .
\end{equation}
The resulting increase of $t_j$ during the RG flow implies that the
system approaches the resonant Andreev reflection fixed point.

However, this scenario gets modified by the charging
term at energy scales $\omega<E_c$, where the zero mode $\Phi_0$ 
is governed by a nearly 'free' action corresponding to
 $G^\pm_0(\omega)\sim 1/\omega^2$ in Eq.~(\ref{epsdef}). 
Integration over the fast zero mode, see \ref{appa}, now generates a 
linear 'confinement' potential between tunneling operators
sitting on the same branch $s=\pm$ of the Keldysh contour,
\begin{equation}\label{had1}
\langle {\cal O}_{j,s}^+(t) {\cal O}_{k,s}^-(t')\rangle^{}_0
\simeq  e^{-\frac{2E_c}{\pi} |t-t'|}  \ {\cal O}_{jk,s} (t) ,
\quad {\cal O}_{jk,s} (t) = e^{i\Phi_{j,s}(t)} e^{-i\Phi_{k,s}(t)},
\end{equation}
which binds particles with flavor $j$ and antiparticles 
with flavor $k\ne j$ together; for $k=j$,
only inconsequential particle-antiparticle annihilation events occur.
Notice that only the $\alpha$ part of $\Phi$ in Eq.~(\ref{alphadef}),
which is orthogonal to the zero mode, determines the ${\cal O}_{jk}$
operators. For low energies, $\omega\lesssim E_c$, the physically relevant
 degrees of freedom then correspond to the 
composite objects ('dipoles') described by  
${\cal O}_{jk}$ -- within our high-energy physics analogy,
these are quark-antiquark pairs ('mesons').  This indicates
that the effective phase action should describe
an interacting \textit{dipole} gas.
The dipoles have symmetric coupling strengths $\lambda_{jk}=
\lambda_{kj}>0$, and therefore $S_t$ will be effectively given
by $\cos$-terms; we put $\lambda_{jj}=0$ since particle-antiparticle
annihilation processes give no dynamical contribution. 
Written again in terms of the phase fields $\Phi=(\Phi_0,\alpha)$,
with $S_0$ defined in Eq.~(\ref{eq:7}),  the low-energy 
Keldysh phase action follows as
\begin{eqnarray} \label{actionstrong}
S[\Phi] &=& S_0[\Phi_0,\alpha] + S_t[\alpha], \\ \nonumber
S_t &=& \sum_{j,k} 
\sum_{s=\pm} s\lambda_{jk} \int dt \cos\left(\Phi_{j,s}-\Phi_{k,s} \right),
\end{eqnarray}
which describes the physics of our system on energy scales $\omega<E_c$.
In physical terms, the $\lambda_{jk}$ describe the amplitude
for processes where a particle is transferred from lead $j$ 
to lead $k$ (or back), with virtual occupation of the dot during 
a timespan of order $E_c^{-1}$. Within our low-energy approach,
this corresponds to instantaneous particle transfer,
dubbed 'teleportation' in Ref.~\cite{fu}.  The 'bare' $\lambda_{jk}$,
defined at the high-energy cutoff scale $\omega=E_c$ of the 
effective action (\ref{actionstrong}), are positive and 
may be estimated as \cite{cmj}
\begin{equation}\label{lam1}
\lambda_{jk}^{(1)} \approx \frac{ t_j^{(1)} t_k^{(1)}}{E_c}\propto E_c^{-3+1/g},
\end{equation}
where a factor $E_c^{-1}$ comes from the time integration over the 
particle-antiparticle separation and the $t_j^{(1)}$ are
specified in Eq.~(\ref{t11}). 

\subsection{Two-channel $\mathrm{SO}(M)$ Kondo effect} \label{sec3b}

We now show that Eq.~(\ref{actionstrong}) naturally describes a 
variant of the two-channel Kondo model with $\mathrm{SO}(M)$ as the 
underlying symmetry group. This connection to Kondo physics
has first been drawn in Ref.~\cite{beri1}. 
In the lead non-interacting limit, $g=1$, the
analogies to the Kondo model can be conveniently exposed in a refermionized
language. Mainly for pedagogical purposes, we briefly discuss this
fermion representation now before returning to the analysis of the
bosonized action for arbitrary $g$. 
Following standard procedures \cite{gogolin}, we represent the fermions 
propagating in the now non-interacting $j$th lead in terms of the
auxiliary right-moving fermion field $\psi_j(x)$, where 'unfolding'
of the semi-infinite wire to an 
infinite chiral wave guide is understood. The inter-wire coupling
introduced by the dot can be represented by refermionization, 
i.e., by writing $e^{i\Phi_j} = \sqrt{a} \ \eta_j \psi_j(0)$.
Notice that the Majoranas $\eta_j$ 
are not identical to the Klein-Majorana factors of
the native model. Likewise, the effective fermions $\psi_j(x)$ differ 
from the original wire fermions.  
The effective fermion Hamiltonian equivalent to the boson  
representation in Eq.~(\ref{actionstrong}) then reads as
\begin{equation}\label{referm}
H_f = -i v \int_{-\infty}^\infty dx \sum_{j=1}^M 
\psi_j^\dagger (x) \partial_x \psi_j^{}(x) 
+ a \sum_{j\ne k} \lambda_{jk}\eta_k\eta_j \psi_j^\dagger(0) \psi_k^{}(0).
\end{equation}
For this model, the coupling constants $\lambda_{jk}$ 
flow under renormalization according to the one-loop RG 
equations \cite{beri1}
\begin{equation} \label{eq:RGnonint}
  \frac{d\lambda_{jk}}{d\ln b}=  
\frac{\kappa}{E_c} \sum_{m\neq (j,k)} \lambda_{jm}\lambda_{mk},
\end{equation}
where $\kappa=\mathcal{O}(1)$ is a non-universal constant. $E_c$
appears as a high-energy cutoff marking the validity limit of
the action (\ref{actionstrong}), and hence of 
the refermionized model (\ref{referm}). For our present 
configuration of initially positive couplings, these equations
predict a flow towards an isotropic configuration,
$\lambda_{jk}\to \lambda (1-\delta_{jk})$, where  
$\lambda$ grows according to
\begin{equation}\label{standardkondo}
 \frac {d\lambda}{ d\ln b} = \frac {\kappa (M-2)}{ E_c} \lambda^2.
\end{equation}
The effective Hamiltonian thus flows towards an \textit{isotropic} limit,
\begin{equation}\label{refermIso}
H_f = -i v \int dx \sum_{j} 
\psi_j^\dagger \partial_x \psi_j
+ J \sum_{j\ne k} \eta_k\eta_j \psi_j^\dagger(0) \psi_k^{}(0),
\end{equation}
with positive coupling $J=a\lambda$.  The bilinears $A_{jk}
\equiv \eta_j \eta_k$ appearing in $H_f$ define an 
$\mathrm{so}(M)$ algebra. To expose the symmetry of the model in its
most obvious form, we pass to a real Majorana basis for
each lead channel, $\psi = \mu + i \nu$ and $\psi^\dagger = \mu - i \nu$, 
whereupon we obtain
\begin{equation}\label{refermIsoReal}
H_f = -i v \int dx \ \mu^T \partial_x \mu - J \mu^T(0) \hat A \mu(0)+
(\mu\leftrightarrow \nu)
\end{equation}
with $\hat A = \{A_{jk}\}$. This defines a variant of the two-channel
$(\mu,\nu)$ Kondo model with symmetry group $\mathrm{SO}(M)$.

\subsection{Scaling equations and Kondo temperature}

We now return to the Keldysh phase action (\ref{actionstrong}) and allow
for $g\le 1$ again.  The RG equations generalizing 
Eq.~(\ref{eq:RGnonint}) may be obtained by standard Coulomb gas
energy-shell integration, or by using
the operator product expansion. The result is \cite{cmj,beri2}
\begin{equation}\label{rg}
\frac{d\lambda_{jk}}{d\ln b}= -\gamma \lambda_{jk}
+ \frac{\kappa}{E_c} \sum_{m\neq (j,k)} \lambda_{jm}\lambda_{mk},
\end{equation}
where $\gamma\equiv g^{-1}-1 >0$.  The first term
reflects the well-known power-law suppression 
of the tunneling density of states for Luttinger liquids \cite{gogolin}, and
leads to a suppression of the $\lambda_{jk}$ under the RG flow.   
For $M>2$ lead channels, the Kondo-like second contribution opposes 
this suppression.  As a result of this competition, an isotropic
intermediate fixed point emerges, 
$\lambda_{jk}=\lambda^* (1-\delta_{jk})$, where
\begin{equation}\label{fp}
\lambda^*= \frac{\gamma}{\kappa(M-2)} E_c.
\end{equation}
Defining $\lambda_{jk} = \lambda^*(1-\delta_{jk})+\mu_{jk}$, the RG flow 
in the vicinity of this fixed point is described by the linearized equations
\begin{equation}\label{rglin}
  {d\mu_{jk}\over d\ln b} = {\gamma\over M-2} \left(-M\mu_{jk}+
    (1-\delta_{jk})\sum_{m=1}^M (\mu_{jm}+\mu_{mk})\right).
\end{equation}
As detailed in \ref{sec:scaling-dimensions}, the solution approaches
the \textit{isotropic} configuration
\begin{equation} \label{eq:2}
  \mu_{jk}\sim \left\langle \mu^{(1)} \right\rangle_{\rm av} 
(1-\delta_{jk}) b^\gamma,
\end{equation}
where the average of coupling constants over all channel indices is denoted by
$\langle \mu\rangle_{\rm av} = {1\over M(M-1)}\sum_{jk} \mu_{jk}$,
and $\mu^{(1)}$ defines the 'bare' couplings
according to Eq.~(\ref{lam1}).  Equation (\ref{eq:2}) shows that (i) 
anisotropic deviations in the $\lambda_{jk}$ correspond
to irrelevant scaling fields,  vanishing during the RG flow
with the non-universal scaling dimensions specified in 
\ref{sec:scaling-dimensions}, and (ii) the fixed point
$\lambda^\ast$ in Eq.~(\ref{fp}) is \textit{unstable}. 
Depending on the average value of the
initial deviation off the critical configuration, the flow is either
to weak coupling (for $\langle \mu^{(1)}\rangle_{\rm av} <0$), 
or towards strong
coupling ($\langle \mu^{(1)}\rangle_{\rm av} >0$). In either case, an
$\mathrm{SO}(M)$-symmetric configuration will be approached.  

To explore what happens in the strong coupling regime,  let us
consider 'bare' couplings with 
$\langle\lambda^{(1)}\rangle_{\rm av} >\lambda^*$.  
Neglecting both the RG-irrelevant anisotropic contributions and the
now inessential term linear in $\lambda$, 
Eq.~(\ref{rg}) simplifies to the standard Kondo form 
(\ref{standardkondo}).  With the 'Kondo temperature' defined by
\begin{equation}\label{tk}
T_K\approx E_c \exp\left(- \frac{1}{\kappa (M-2)}
\frac{E_c}{ \langle\lambda^{(1)}\rangle_{\rm av} }
\right),
\end{equation}
the resulting RG flow diverges at $\omega\sim T_K$.
Clearly, the perturbative RG analysis then ceases to be valid.  
The physics on even lower energy scales is best discussed 
by switching to a dual action, as we discuss next.

\subsection{Dual Keldysh phase action: Below $T_K$}

Assuming $\langle\lambda^{(1)}\rangle_{\rm av} >\lambda^*$ and
very low energy scales $\omega\lesssim T_K$, 
the coupling $\lambda$ effectively approaches the strong-coupling 
limit, where the fields $\Phi_j$ are confined near the minima
of $S_t$ in Eq.~(\ref{actionstrong}). 
The dominant excitations of $\Phi=(\Phi_0,\alpha)$
are occasional tunneling events between 
neighboring minima (in a slight abuse of notation referred
to as 'instantons'), where $\Phi_{j,s} \to \Phi_{j,s} \pm 2\pi$. 
Noting that $\Phi_0$ does not enter the tunneling action 
in Eq.~(\ref{actionstrong}), $S_t=S_t[\alpha]$,
we now perform a Hubbard-Stratonovich transformation
to dual fields $(\Theta_0,\beta)$, corresponding to the 
conjugate lead boson fields $\{ \theta_j(x=0) \}$. 
We thus represent $S_0[\Phi_0,\alpha]$ in Eq.~(\ref{eq:7}) as 
\begin{eqnarray} \label{eact} 
S_0 &=& {1\over 2} \int {d\omega\over 2\pi}
\left[\Theta^T_{0}(-\omega) \tilde G_0^{-1}(\omega) \Theta_0(\omega)-
\beta^T (-\omega) \tilde G^{-1}(\omega)\otimes \Delta \beta(\omega) \right] \\ 
\nonumber & +&
i\int {d\omega\over 2\pi} \frac{2\omega}{\pi}
\left[\Phi^T_{0}(-\omega) \tau_1\Theta_{0}(\omega)+
\alpha^T(-\omega) \tau_1\otimes \Delta\beta(\omega) \right],
\end{eqnarray}
with the $M-1$ time-dependent 'discrete' variables
$\beta_{i}=(\beta_{i,c},\beta_{i,q})^T$, and $\beta_0\equiv 0$.
Here we used the reciprocity relation $G(\omega) = -(\pi/2\omega)^{2}
\tau_1 \tilde G^{-1}(\omega) \tau_1$, where $\tilde G=g^2 G$ 
differs from $G$ only by the parameter exchange $g\to 1/g$ 
and the Pauli matrix $\tau_1$ acts in Keldysh space. 
Similar relations hold for $\tilde G_0$.
The linear couplings in the second line of the
transformed action indicate that the variables $( \Phi_0,\Theta_0)$ 
and $(\alpha,\Delta \beta)$ indeed form canonical
pairs.  

The integration over the unrestricted zero mode $\Phi_0$ now
generates the constraint $\Theta_0=0$, which in physical terms
implies current conservation at the dot. 
Turning to the nonlinear variables $\alpha$, the least costly
excitations correspond to $m$-instanton configurations 
defined by a sequence of $2\pi$-steps occurring at times $t_a$, 
$\Phi_{i_a,s_a}\to \Phi_{i_a,s_a}+2\pi\sigma_{a}$, 
with Keldysh indices $s_a=\pm$ and jump directions $\sigma_a=\pm$
($a=1,\ldots,m$).  Taking into account the constraint 
$\Theta_0=0$, the Fourier representation of this multi-instanton 
profile follows from the equation
\begin{equation}\label{latt}
i\omega \Delta \alpha(\omega) = 2\pi \sum_{a=1}^m e^{i\omega t_a}
\sigma_a g_{s_a}\otimes (f_{i_a}-f_{i_a-1}),
\end{equation}
with the Keldysh vectors $g_\pm={1\over 2}(1,\pm 1)^T$,
the standard unit vectors $f_i$ in $(M-1)$-dimensional space,
and $f_0\equiv 0$.  Geometrically, the solutions to Eq.~(\ref{latt})
correspond to lattice vectors of a hyper-triangular lattice embedded into 
the $(M-1)$-dimensional hyperplane perpendicular to the vector $e_0$.  
Substituting Eq.~(\ref{latt}) into the action (\ref{eact}), and
denoting the tunneling action for a single instanton by 
$iS_\mathrm{inst}$ \cite{foot2}, we obtain the multi-instanton action 
as $imS_{\rm inst}+S^{(m)}$ with
\[
S^{(m)}=- {1\over 2} \int {d\omega\over 2\pi}
\beta^T (-\omega) \tilde G^{-1}(\omega)\otimes \Delta
   \beta(\omega) -\sum_{a} \sigma_a s_a \left(\beta_{i_a,s_a}-
\beta_{i_a-1,s_a}\right)(t_a).
\]
Integrating over the instanton times $t_a$, summing over 
$i_a=1,\ldots,M-1$ and the indices $\sigma_a,s_a=\pm$, 
and taking into account all orders $m$, we finally arrive at the 
\textit{dual Keldysh phase action},
\begin{eqnarray} \label{eq:1}
S[\beta]& =&-\frac12 \int \frac{d\omega}{2\pi}
\beta^T(-\omega) \tilde G^{-1}(\omega)\otimes \Delta 
\beta(\omega) \\ \nonumber &+&
 y \sum_{s=\pm} s \sum_{j=1}^{M-1}\int dt \
\cos \left(\beta_{j,s}-\beta_{j-1,s}\right),
\end{eqnarray}
where the coupling constant $y\sim e^{-S_\mathrm{inst}}$ vanishes 
for $\lambda\to\infty$.   We have also derived the 
action (\ref{eq:1}) by using a Villain approximation of 
the $\cos$-terms in Eq.~(\ref{actionstrong}),
similar to the approach taken in Ref.~\cite{fidkowski} for the single
lead ($M=1$) case.

The scaling dimension $1-\Delta_M$ 
of the nonlinear perturbation $\sim y$ in Eq.~(\ref{eq:1})
now follows from the auxiliary relations $(\Delta^{-1})_{ii'} =-(M-i_>)i_</M$,
where $i_{</>}$ is the smaller/larger of the indices $i$ and
$i'$, and 
\begin{equation} \label{eq:5}
\left\langle [\beta_{i}-\beta_{j}]_s(-\omega)
\, [\beta_{i}-\beta_{j}]_{s'}(\omega)\right\rangle_{y=0}=
\tilde G_{ss'}(\omega) \frac{(M-|i-j|)|i-j|}{M}.
\end{equation}
First-order renormalized perturbation theory \cite{gogolin} 
then yields
\begin{equation}\label{deltam}
\Delta_M = 2g \left(1-\frac{1}{M}\right),
\end{equation}
see also Refs.~\cite{beri2,kane,yi}.   For 
weakly repulsive interactions, $\frac{M}{2(M-1)}<g\leq 1$,
the scaling dimension $1-\Delta_M$ is negative and hence describes a
RG-irrelevant perturbation.  At higher orders in perturbation 
theory, additional operators $\sim \cos(\beta_i-\beta_j)$ 
with $|i-j|>1$ may be generated. However, Eq.~(\ref{eq:5}) implies 
that these are even more irrelevant than the perturbation in Eq.~(\ref{eq:1}). 
Overall, the analysis above demonstrates the stability of the 
strong-coupling $\mathrm{SO}(M)$ Kondo fixed point $\lambda\to \infty$.  

\subsection{Discussion}

\begin{figure}[t]
\begin{center}
\includegraphics[width=0.7\textwidth]{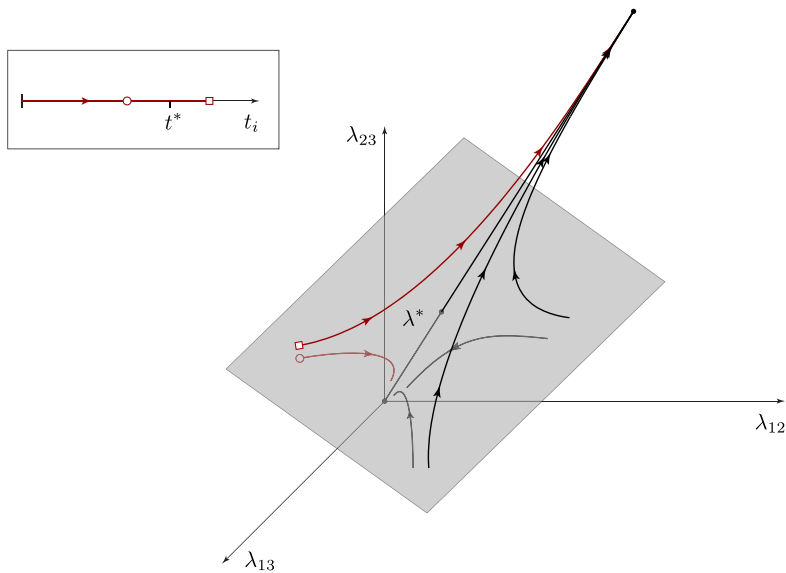}
\end{center}
\caption{\label{fig2} Illustration of the RG flow of coupling
constants for $M=3$ leads. At high energy scales
$\omega\gtrsim E_c$ (inset), tunneling couplings effectively
increase as $t_i \sim \omega^{-1+{1\over 2g}}$ during the
first stage of the RG flow.  Their terminal value, reached at $\omega\sim E_c$, 
then determines the initial 'bare' dipole couplings,
$\lambda^{(1)}_{jk}\propto E_c^{-3+1/g}$, see  Eq.~(\ref{lam1}).
These dipole couplings enter the effective low-energy action 
(\ref{actionstrong}), and constitute the scaling variables
during the second stage of the RG flow (where $\omega\lesssim E_c$). 
Depending on whether the average value $\langle \lambda^{(1)}\rangle_{\rm av}$ 
is larger or smaller than the repulsive fixed point $\lambda^\ast\propto E_c$,
the system now flows either to strong coupling ($\lambda\to \infty$), or to 
the decoupled fixed point $\lambda=0$; cf.~the trajectories 
starting with {\scriptsize $\square$} and $\circ$ symbols, resp. 
Inter-channel deviations in the coupling strengths 
are irrelevant and scale to zero with non-universal scaling dimensions.}
\end{figure}

Let us now summarize the full picture emerging for the scaling of the
coupling constants all the way from high energy scales (comparable to
the bandwidth $\Lambda$ set by the proximity gap) down to the
actually probed scale set by temperature or applied voltages;
for illustration, see Fig.~\ref{fig2}. 
Lowering the energy scale from $\omega=\Lambda$, the direct single-particle
tunneling couplings $t_j$ grow independently.  This growth
will stop at the energy scale $\omega\sim E_c$, where confinement sets in.
The weaker the charging energy $E_c$, the larger the coupling constants 
may become before this flow stops, see  Eq.~(\ref{t11}). 
For $\omega\lesssim E_c$, the theory is instead governed by a
system of 'dipoles' coupled at strength $\lambda_{jk}$, with the 'bare'
values $\lambda^{(1)}_{jk}$ in Eq.~(\ref{lam1}). 
Everything now depends on whether the 'bare' coupling 
strength averaged over all configurations,  
$\langle\lambda^{(1)}\rangle_{\rm av}$, is larger or smaller 
than the unstable isotropic fixed point $\lambda^\ast$ in Eq.~(\ref{fp}).
For $\langle\lambda^{(1)}\rangle_{\rm av} >\lambda^*$, the 
RG flow proceeds towards a
strong-coupling $\mathrm{SO}(M)$-symmetric fixed point,
$\lambda_{j\ne k}=\lambda\to \infty$. 
Deviations in the strength between different $\lambda_{jk}$ scale
to zero, where the details of the flow are 
discussed in \ref{sec:scaling-dimensions}. 
Note that in conventional multi-channel Kondo proposals,
anisotropy is a relevant perturbation and easily destabilizes
the Kondo fixed point \cite{gogolin}.  In contrast, 
the present system is robust in that it flows towards an 
isotropic configuration. 
If the charging energy is too large for $\langle\lambda^{(1)}\rangle_{\rm av}$ 
to reach $\lambda^\ast$, the flow will be towards an
equally isotropic configuration with $\lambda\to 0$ \cite{cmj}. 
In this limit, we recover the conventional Luttinger liquid junction behavior
\cite{oshikawa,nayak,chen,alex}, where all leads  effectively decouple
from the dot at very low energies. 

\section{Transport observables}\label{sec4}

In this section, we address the differential 
conductance tensor, $G_{jk}$, defined
in Eq.~(\ref{conddef}), and the
shot noise tensor, $S_{jk}=S_{jk}(\omega=0, T=0)$, defined 
in Eq.~(\ref{noisedef}).  Within our phase action approach, $Z[\chi]$ 
has been represented as Keldysh functional integral over the phase fields
$\Phi=(\Phi_0,\alpha)$, or the dual $\beta$ variables. 
For large $E_c$ and $g<1$, the system flows towards the decoupled fixed point
$\lambda=0$, where perturbative expansion in $\lambda_{jk}$ with 
the action (\ref{actionstrong}) yields the low-energy
dependence of all transport observables.  
This decoupled fixed point has been studied in depth before 
\cite{gogolin,chen} and implies a Luttinger power-law suppression of the 
linear conductance, $G_{jk}\sim T^{2/g-2}$, at low temperatures.
The shot noise near the decoupled fixed point has also been
analyzed \cite{trauzettel}. 

We here focus on the case of intermediate charging 
energy $E_c$ with $M>2$ leads, where the RG flow is towards the 
strong-coupling Kondo fixed point as long as $\Delta_M>1$, with $\Delta_M$ 
in Eq.~(\ref{deltam}). 
Using the energy scale $\Omega={\rm max}(|\mu_j-\mu_k|,T)$,
we may now distinguish three different regimes.
First, in the high energy regime, $\Omega>E_c$,
the charging energy does not significantly affect the
noninteracting resonant Andreev reflection scenario.
The Keldysh phase action is then given by 
Eq.~(\ref{eq:7}) with $G_0\to G$, and it is straightforward
to derive the temperature dependence of the linear conductance,
 $G_{jk}\sim T^{-2+1/g}$.  Putting $g=1$,  
the well known $1/T$ scaling of the  
zero bias anomaly peak conductance at high temperatures
is recovered \cite{alicea}.  
Similarly, the shot noise here corresponds to the Fano factor $F=2$ found 
in the resonant Andreev reflection regime \cite{golub}.   

Proceeding to lower energies, $\Omega<E_c$, the charging energy implies
dipole formation as described by the action $S[\Phi]=S_0+S_t$ in
Eq.~(\ref{actionstrong}).  The chemical potentials and
the counting fields can be included by shifting 
$\Phi_j(t)\to \Phi_j(t)+(\tilde \mu_j t, \tilde\chi_j(t)/2)^T$ in $S_t$.
Gauge invariance implies that these appear only through the quantities 
\begin{equation}\label{tildemu}
\tilde\mu_j = \mu_j -\frac{1}{M}\sum_{k=1}^M \mu_k,\quad
\tilde\chi_j = \chi_j -\frac{1}{M}\sum_{k=1}^M \chi_k.
\end{equation}
The regime $T_K<\Omega<E_c$ could then be analyzed by perturbation
theory in the $\lambda_{jk}$. 

However, we here only discuss the most interesting 
low-energy regime, $\Omega<T_K$, where the dual Keldysh 
action $S[\beta]$ in Eq.~(\ref{eq:1}) applies. 
The chemical potentials and counting fields
then yield the additional action piece
\begin{equation} \label{deltaS}
S_V[\beta] = - \frac{2}{\pi} \sum_{j=1}^{M} \int dt \
V^T_j \tau_1 \left( \beta_j- \beta_{j-1} \right),  \quad
V_j(t)= \left( \begin{array}{c}\tilde\mu_j \\
 \frac12 \dot{\tilde\chi}_j \end{array}\right),
\end{equation}
with the Pauli matrix $\tau_1$ in Keldysh space.
By using $\tilde\mu_j$ and $\tilde\chi_j$,  
current conservation is automatically maintained, and thus the conductance
sum rule $\sum_{j=1}^M G_{jk}=0$ always holds. 
Since the nonlinear perturbation $\sim y$ in Eq.~(\ref{eq:1}) is RG-irrelevant,
the transport observables for $\Omega<T_K$ follow by expanding
$\ln Z[\chi]=\sum_{n=0}^\infty \ln Z^{(n)}[\chi]$ in powers of $y$, 
where we report only on the lowest two nontrivial orders ($n=0,2$).  
The unitary limit behavior follows by putting $y=0$ in $S[\beta]$.
Performing the remaining Gaussian field integration over $\beta$, we find
\begin{equation}\label{lowTfunc}
\ln Z^{(0)}[\chi] = - \frac{2i}{\pi^2} \int \frac{d\omega}{2\pi}
V^T(-\omega) \tau_1 \tilde G(\omega) \tau_1  V(\omega).
\end{equation}
Some algebra gives for the second-order contribution the result
\begin{eqnarray}\label{z2}
\ln Z^{(2)}[\chi] &=& \frac{y^2}{2} \sum_{j=1}^M \sum_{\sigma=\pm}
\int dt_1 dt_2 e^{-2J(t_1-t_2)-2i\sigma g \tilde\mu_j(t_1-t_2) } 
\times \\ \nonumber
&\times& \prod_{r=1,2} \sin\left[ K(t_1-t_2) + 
g\sigma \tilde\chi_j(t_r)\right],
\end{eqnarray}
where the lead 'bath' correlation function is
given by
\begin{equation}\label{bathcor}
J(t)-iK(t) = \Delta_M \int_0^{E_c} \frac{d\omega}{\omega} 
\left\{ [1-\cos(\omega t)] \coth(\omega/2T) - i \sin(\omega t)\right\}.
\end{equation}

The linear conductance tensor then follows from Eq.~(\ref{conddef}),
see also Ref.~\cite{beri2},
\begin{equation}\label{unitary}
G_{jk}(T\ll T_K) = \frac{2g e^2}{h} \left( \delta_{jk} - \frac{1}{M} \right) 
\left [ 1- c_0 (T/T_K)^{2\Delta_M-2} + \cdots \right],
\end{equation}
where $c_0={\cal O}(1)$ and
\begin{equation}
\label{eq:6}
T_K\equiv  \left(\frac{ \Gamma(2\Delta_M) E^2_c} 
{2\pi g^2y^2}\right)^{1/2(\Delta_M-1)} \frac{E_c}{2g},
\end{equation}
with $\Gamma$ denoting the Gamma function. 
Equation (\ref{eq:6}) \textit{defines} the
Kondo temperature from the perspective of the strong coupling
theory. In Eq.~(\ref{tk}), we had identified $T_{K}$ as the
\textit{low} energy scale where the coupling constants of the 
weak coupling theory diverge. At lower energies, we are 
operating in the realm of the
dual strong coupling theory discussed presently. The validity regime
of the latter is limited by a \textit{high} energy scale
$\sim T_{K}$, where the corrections due to infrared irrelevant
nonlinearities remain strong enough to produce $\mathcal{O}(1)$
corrections to the asymptotic Gaussian fixed point theory. Our
perturbative analysis identifies this scale as in Eq.~(\ref{eq:6}),
which may be regarded as a definition of $T_{K}$ in terms of the
dual coupling constant.  

Equation (\ref{unitary}) describes an 
isotropically hybridized multiterminal junction.
Following standard arguments \cite{lutchyn,maslov,safi,pono},
when the one-dimensional nanowire 'leads' are eventually
connected to wide bulk reservoirs, the prefactor $g$
in Eq.~(\ref{unitary}) is replaced by the
Fermi liquid value of the reservoirs, $g\to 1$.
The $T^{2\Delta_M-2}$ power-law corrections 
to the unitary limit should be contrasted to the corresponding
$\sqrt{T}$ temperature dependence for the 
two-channel $\mathrm{SU}(2)$ Kondo case \cite{twochannel}. 

In the zero temperature limit, Eq.~(\ref{bathcor}) yields
$J(t)\simeq \Delta_M \ln(E_c|t|)$ and
$K(t)\simeq \frac{\pi}{2} \Delta_M \ {\rm sgn}(t)$.
It is then straightforward to establish that the currents, 
\begin{equation}
I_j= \frac{2ge^2}{h} \sum_{k=1}^M \left (\delta_{jk}-\frac{1}{M}\right) 
\mu_k + I^{(2)}_j + \cdots,
\end{equation} 
receive the 'backscattering' corrections 
\begin{equation}\label{bsc}
I_j^{(2)} = - \frac{e}{\hbar} \sum_{k=1}^M 
\left|\frac{\tilde\mu_k}{T_K}\right|^{2\Delta_M-2}
\left(\delta_{jk}-\frac{1}{M}\right) \tilde\mu_k.
\end{equation}
Turning to the $T=0$ shot noise tensor, we find 
that a finite contribution may arise only in order $y^2$.
This shot noise suppression in the unitary limit is a direct consequence of 
the 'free' zero mode dynamics. Note that also  
the cross-correlations between different terminals are suppressed,
despite of the current partitioning implied by Eq.~(\ref{unitary}).
Equation (\ref{z2}) yields the shot noise tensor
\begin{equation}
S_{jk} = -\frac{2g e^2}{\hbar} 
\sum_{l=1}^M \left(\delta_{jl}-\frac{1}{M}\right)
\left(\delta_{kl}-\frac{1}{M}\right) \left|\frac{\tilde\mu_l}{T_K}
\right|^{2\Delta_M-2} |\tilde\mu_l|.
\end{equation}
To define the Fano factors in this multiterminal 
setting, it is customary \cite{rosa} to compare $S_{jk}$ to the backscattered
currents (\ref{bsc}). Writing $\tilde\mu_j\sim V$ with an 
overall 'voltage' scale, we observe that the Fano factors are again
\textit{universal} (independent of $V$ or
$T_K$).  Taking, for instance, $\mu_1=V$ and $\mu_{j>1}=-V/(M-1)$,
the Fano factor pertaining to the first lead is
\[
 F_1= \frac{S_{11}}{e I^{(2)}_1} 
= \frac{2g}{M} \frac{(M-1)^{2\Delta_M}+1}{(M-1)^{2\Delta_M-1}+1}.
\]
For $M\gg 1$, we recover the effectively noninteracting Fano
factor $F=2$ predicted by the resonant Andreev reflection picture \cite{golub}.
Shot noise measurements could thus probe the noninteger
scaling dimensions $\Delta_M$ associated with 
non-Fermi liquid behavior in this two-channel 
$\mathrm{SO}(M)$ Kondo problem.

\section{Concluding remarks}\label{sec5}

In this paper we have analyzed a multiterminal
Coulomb-Majorana junction, where the junction is
formed by a mesoscopic superconductor containing 
Majorana bound states due to the presence of helical nanowires. 
For $M$ attached leads, a  two-channel Kondo model with symmetry group
$\mathrm{SO}(M)$ emerges when the charging
energy of the 'dot' is finite.  Salient features of this Kondo effect include
dynamically generated universality --- no fine tuning of coupling
constants is required to establish the underlying $\mathrm{SO}(M)$
symmetry ---, and non-Fermi liquid scaling in the vicinity of the
strong coupling fixed point. It stands to reason that this scaling
might become observable by transport measurements.

We close by re-emphasizing two assumptions crucial to the physics
discussed here. 
First, the relevant energy scales (e.g., temperature, applied voltages,
or the charging energy) should be below the superconducting gap to avoid
quasiparticle excitations. Second,
direct tunneling processes between Majorana bound states are assumed absent.
This issue is probably important for presently discussed
implementations, where the typical MBS size is
believed to be of the order of several 100 nm.  
Direct tunneling is an RG-relevant perturbation, like the
magnetic Zeeman field in the usual Kondo problem, and is expected
to strongly affect the Kondo physics reported here.  Since 
tunneling spoils the usefulness of our Klein-Majorana fusion 
trick, a modified theoretical approach would also be necessary
to describe this situation.
We hope that future work will address this challenge, 
as well as the experimental realization of this proposal.

\ack
We wish to thank B. B{\'e}ri, L. Fu and A. Tsvelik for helpful discussions.
We acknowledge support by the DFG within the research networks SFB TR 12
and SPP 1666.

\appendix
\section{Winding number summation}\label{appa}

Here we address the summation over the
integer winding numbers $W$ appearing
in Eq.~(\ref{sc2}). For simplicity, we 
switch to the imaginary time ($\tau$) version of the theory.
Taking into account the tunneling action (\ref{tunnel}) and
integrating over the bulk lead modes, the partition sum
has the following functional integral representation 
over the boundary boson fields $\Phi_j(\tau)$ and the condensate
phase field $\varphi(\tau)$:
\begin{eqnarray*}
Z &=& \sum_{W=-\infty}^\infty e^{2\pi i n_g W}
\int D{\varphi} e^{- \frac{1}{4E_c}\int d\tau\dot\varphi^2}  \int D\Phi
e^{ -S_l[\Phi] - S_{t}[\Phi,\varphi] },\\
S_l &=& \frac{Tg}{2\pi} \sum_{j=1}^M 
\sum_\omega |\omega| |\Phi_j(\omega)|^2, \quad
S_t = \sum_j t_j \int d\tau \sin(\Phi_j+\varphi),
\end{eqnarray*}
with $\varphi(\tau+1/T)=\varphi(\tau)+2\pi W$ and
bosonic Matsubara frequencies $\omega$.
Writing $\varphi(\tau)=
\tilde \varphi(\tau)+2\pi W T\tau$, shifting $\Phi_j\to \Phi_j-
\tilde\varphi$, and performing the Gaussian functional integral
over $\tilde \varphi$, we obtain 
\begin{eqnarray*}
Z&=&\sum_W e^{2\pi i n_g W - \frac{\pi^2 T}{E_c} W^2 } Z^{(W)}, \quad
Z^{(W)}= \int D{\Phi} e^{-S_l[\Phi]- S^{(W)}_t[\Phi]}, \\
S_l&=& \frac{Tg}{2\pi} \sum_\omega |\omega| \left( 
\frac{1}{1+\epsilon/|\omega|} |\Phi_0(\omega)|^2 -
\alpha^T(-\omega) \Delta \alpha (\omega) \right), \\
S_t^{(W)}&=& \sum_j t_j \int d\tau \sin(\Phi_j+2\pi W T\tau),
\end{eqnarray*}
with $\epsilon\sim E_c$ in Eq.~(\ref{epsdef}). In the 
dissipative action $S_l[\Phi=(\Phi_0,\alpha)]$, the
zero mode $\Phi_0= \sum_j \Phi_j/\sqrt{M}$ has been isolated,
where $(\alpha_1,\ldots,\alpha_{M-1})$ is the orthogonal
complement, see Eq.~(\ref{alphadef}). The matrix $\Delta$ 
has been specified in Eq.~(\ref{eq:3}).

\begin{figure}[t]
\begin{center}
\includegraphics[width=0.8\textwidth]{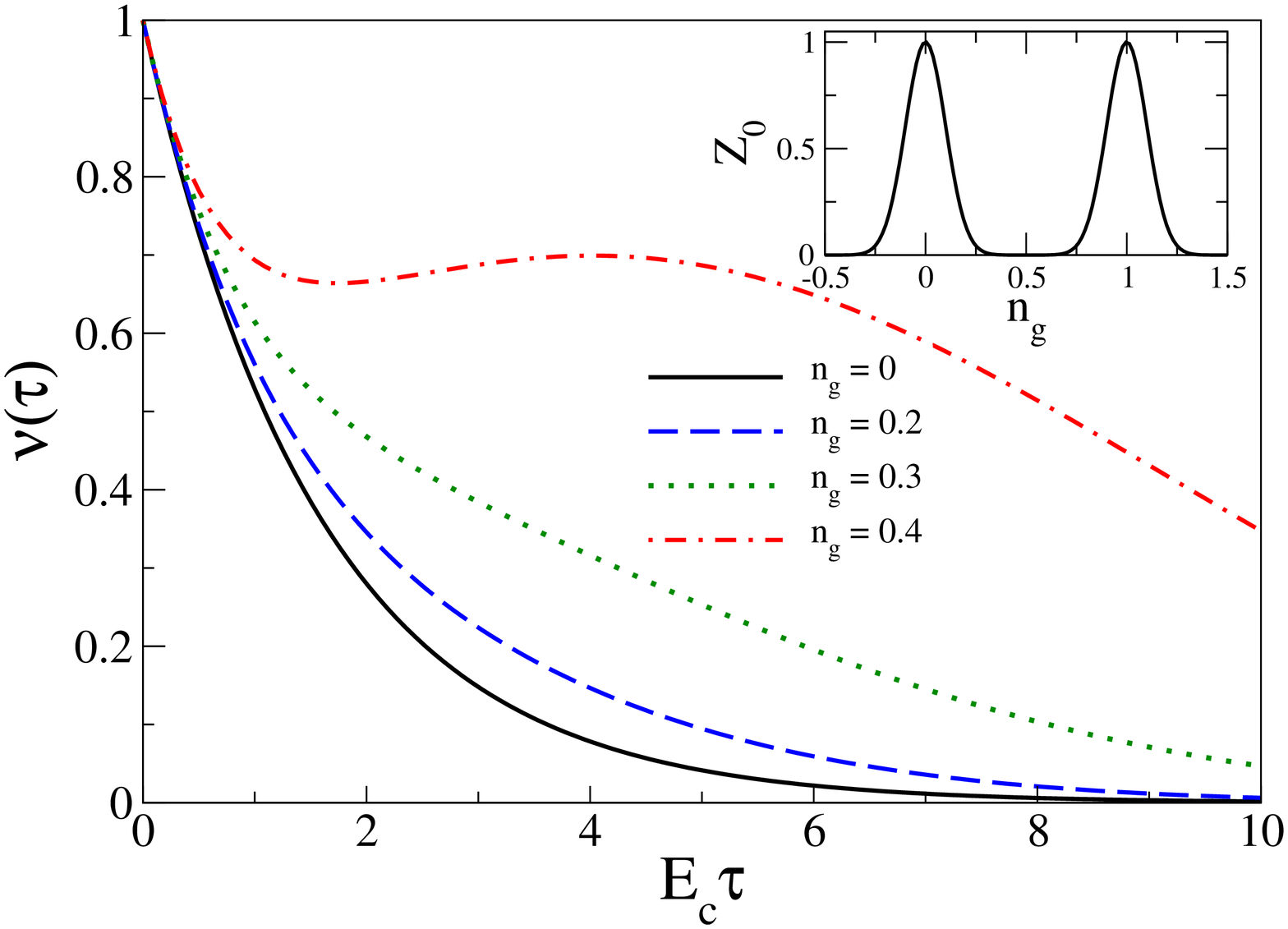}
\end{center}
\caption{\label{fig3}
Main panel: Confinement potential $\nu(\tau)$ defined in Eq.~(\ref{kernW}) 
for $E_c/T=50$ and various values of the gate voltage parameter $n_g$.
Inset: Partition function $Z_0(n_g)$ of the isolated dot as a function
of $n_g$, again for $E_c/T=50$.  }
\end{figure}

We are now ready to integrate over the 'free' zero mode $\Phi_0$, 
which yields $Z^{(W)}=\int D\alpha e^{-S_l[\alpha]-S^{(W)}_t[\alpha]}$
with
\begin{eqnarray*}
S^{(W)}_t &=& -\frac{1}{4} \sum_{j\ne k} t_j t_k 
\int d\tau_1 d\tau_2\  \nu^{(W)}(\tau_1-\tau_2) 
\cos[\Phi_j(\tau_1)- \Phi_k(\tau_2)] ,\\
\nu^{(W)}(\tau)&=& e^{-2E_c|\tau|/\pi} \cos(2\pi W T\tau) .
\end{eqnarray*}
Summation over $W$ writes the partition sum as
$Z=\int D\alpha e^{-S_l[\alpha] -S_t[\alpha]}$, 
where second-order cumulant expansion in the tunneling amplitudes $t_j$ 
 gives $S_t=\langle S_t^{(W)}\rangle_W$. In effect, the kernel $\nu^{(W)}$
is thereby replaced by 
$\nu(\tau)=\langle \nu^{(W)}(\tau)\rangle_W$.  Explicitly, we find
the 'dipole confinement' kernel
\begin{equation}\label{kernW}
\nu(\tau) = e^{-2E_c|\tau|/\pi} \frac{\vartheta(n_g+T\tau,i\pi T/E_c)+
\vartheta(n_g-T\tau,i\pi T/E_c)}{2Z_0(n_g)},
\end{equation}
where $\vartheta$ is the Jacobi theta function,
and $Z_0(n_g)=\vartheta(n_g,i\pi T/E_c)$ is the partition
function of the isolated dot; note that $Z_0(n_g+1)=Z_0(n_g)$.
The kernel (\ref{kernW}) is shown for various values of $n_g$
in Fig.~\ref{fig3}.  
For nearly integer $n_g$, the winding number average has 
little effect on the confinement kernel,  which
is well approximated by retaining only the $W=0$
sector,  $\nu(\tau)\approx \nu^{(W=0)}(\tau)$.
Only when $n_g$ is close to half-integer values, 
dipole formation -- which is induced by an exponential decay of
the kernel $\nu(\tau)$ -- will be disrupted.
We here assume $n_g$ to stay away from half-integer values, 
such that winding number effects play no important role.
In the main text, we then discuss only the $W=0$ sector and approximate
Eq.~(\ref{kernW}) by Eq.~(\ref{had1}).

\section{Scaling dimensions} \label{sec:scaling-dimensions}

In order to solve Eq.~(\ref{rglin}), we introduce the
discrete Fourier representation
\begin{equation*}
  \tilde\mu_{qp} = \sum_{jk} \mu_{jk}\, e^{i(jq+kp)},\qquad \mu_{jk}=
  {1\over M^2}\sum_{qp}
  \tilde\mu_{qp}\, e^{-i(jq+kp)},
\end{equation*}
where $q,p\in\{0,\dots,M-1\} (2\pi/M)\ \mathrm{mod}(2\pi)$, 
and the symmetry $\mu_{jk}=\mu_{kj}$
translates to $\tilde\mu_{qp}=\tilde\mu_{pq}$. 
Using this representation, we obtain
\begin{equation*}
  {d\tilde\mu_{qp}\over d\ln b} ={\gamma M \over
    M-2}\left(-\tilde\mu_{qp}+\delta_{p,0}\tilde\mu_{q0}+
\delta_{q,0}\tilde\mu_{0p}-{1\over
      M}(\tilde\mu_{(p+q)0}+\tilde\mu_{0(p+q)})\right).
\end{equation*}
This implies that
\begin{eqnarray*}
  q,p\not=0&,\qquad  {d\tilde\mu_{qp}\over d\ln b} =-{\gamma \over
    M-2}\left(M\tilde \mu_{qp}+
\tilde\mu_{(p+q)0}+\tilde\mu_{0(p+q)}\right),\cr
q\not=0&,\qquad  {d\tilde\mu_{q0}\over d\ln b} =-{2\gamma  \over
  M-2}\tilde\mu_{q0},\cr
&\qquad {d\tilde\mu_{00}\over d\ln b} =\gamma \tilde\mu_{00}. 
\end{eqnarray*}
According to these equations, (i) the Fourier mode $\tilde\mu_{00}$ grows
as $\tilde\mu_{00} = \tilde\mu^{(1)}_{00} b^{\gamma}$, where 
$\tilde\mu^{(1)}_{qp}$ follows from the 'bare' coupling constants.
 (ii) Generic modes $\tilde\mu_{qp}$ decay with the
dimensions specified above, with the exception (iii) of
modes $\mu_{q,2\pi-q}$.  These modes exhibit the RG scaling 
$\tilde\mu_{q,2\pi-q}\sim -{1\over M-1} \tilde\mu^{(1)}_{00} b^\gamma$.
Substituting this result back into the 
inverse Fourier representation, and using
$\tilde\mu_{00}^{(1)}=\sum_{jk} \mu_{jk}^{(1)}$, we obtain Eq.~(\ref{eq:2}).


\begin{thebibliography}{10}
\bibitem{foot}
While the crossover scale in Eq.~(\ref{eq:7}) is given by $\epsilon$, 
confinement sets in for $|\omega|<E_c$, see Eq.~(\ref{had1}).  
Since $\epsilon\gg E_c$ for $M\gg 1$, 
this may leave room for an intermediate regime. However, this will
not qualitatively affect the results discussed here.
\end{thebibliography}
\end{document}